\documentclass[prb,twocolumn,superscriptaddress,showpacs,amsmath,amssymb]{revtex4-1}

\usepackage{graphicx}
\usepackage{latexsym}
\usepackage{amsmath}
\usepackage{amssymb}
\usepackage{amsfonts}
\usepackage{color}
\usepackage{bm}
\usepackage{verbatim}
\usepackage[]{color}
\definecolor{red}{rgb}{1,0,0}
\usepackage{framed}
\definecolor{shadecolor}{RGB}{222,222,221}
\definecolor{MS-color}{RGB}{128,0,128}

\begin{document}

\title{Magnetization reversal in S/F/S Josephson junctions on a 3D topological insulator.}

 \date{\today}
 
\author{I. V. Bobkova}
\affiliation{Institute of Solid State Physics, Chernogolovka, Moscow
reg., 142432 Russia}
\affiliation{Moscow Institute of Physics and Technology, Dolgoprudny, 141700 Russia}
\affiliation{National Research University Higher School of Economics, Moscow, 101000 Russia}

\author{A. M. Bobkov}
\affiliation{Institute of Solid State Physics, Chernogolovka, Moscow reg., 142432 Russia}

\author{I. R. Rahmonov}
\affiliation{BLTP, Joint Institute for Nuclear Research, Dubna, Moscow Region, 141980, Russia}
\affiliation{Umarov Physical Technical Institute, TAS, Dushanbe, 734063, Tajikistan}

\author{A.A. Mazanik}
\affiliation{Moscow Institute of Physics and Technology, Dolgoprudny, 141700 Russia}
\affiliation{BLTP, Joint Institute for Nuclear Research, Dubna, Moscow Region, 141980, Russia}

\author{K. Sengupta}
\affiliation{School of Physical Sciences, Indian Association for the Cultivation of Science, Jadavpur, Kolkata-700032, India}

\author{Yu. M. Shukrinov}
\affiliation{BLTP, Joint Institute for Nuclear Research, Dubna, Moscow Region, 141980, Russia}
\affiliation{Dubna State University, Dubna,  141980, Russia}

\begin{abstract}
We study a magnetization reversal  by an electric current pulse in a superconductor/insulating ferromagnet/superconductor Josephson junction placed on top of a 3D topological insulator. It is demonstrated that such a system is perspective for low-dissipative spintronics because of the strong spin-momentum locking in the TI surface states. This property provides an ideally strong coupling between the orbital and spin degrees of freedom thus  giving a possibility of efficient reversal of the magnetic moment by current pulse with amplitude lower than the critical current, that results in strongly reduced energy dissipation. The underlying physical mechanism of the reversal is discussed. The influence of the magnetic anisotropy on the controllability of the reversal by the pulse duration is investigated. In addition, a way of a simultaneous electrical detection  of the reversal is proposed. 
\end{abstract}

  \maketitle
 
\section{Introduction}

Superconducting spintronics \cite{Linder2015} is in the focus of active research now in view of necessity of minimizing energy consumption of devices.   In particular, the ability to reverse magnetic moment of an interlayer magnet by the Josephson current has attracted much attention recently \cite{Shukrinov2017,Chudnovsky2016,Cai2010,Guarcello2020,Atanasova2019,Mazanik2020}. For the most part of the proposed setups the key ingredient allowing for the magnetization reversal  is the spin-orbit coupling (SOC) taking place inside the ferromagnet or if the structure is realized as a complex interlayer consisting of the ferromagnet and a heavy metal exhibiting strong SOC. The SOC leads to the appearance of the anomalous ground state phase shift $\varphi_0$ \cite{Buzdin2008,Assouline2019,Mayer2020} in the corresponding Josephson junction. The strong dependence of $\varphi_0$ on the magnetization direction has been exploited in order to establish coupling between the Josephson and magnetic subsystems. Here we investigate the perspectives of a magnetization reversal in Josephson junctions via 3D topological insulator (TI) surface states. Our study is motivated by the fact that due to the property of spin-momentum locking  of the 3D TI surface states \cite{Burkov2010,Culcer2010,Yazev2010,Li2014} this material can be considered as a system, where the ideally strong SOC occurs. 

S/TI/S Josephson junctions have been experimentally realized recently\cite{Veldhorst2012}. On the other hand, at present there is great progress in the  experimental realization of F/TI hybrids. In particular, to introduce the ferromagnetic order into the TI, random doping of transition metal elements, e.g., Cr or V, has been employed \cite{Chang2013,Kou2013,Kou2013_2,Chang2015}. The second option, which has been successfully realized experimentally, is a coupling of a nonmagnetic TI to a
ferromagnetic insulator (FI) to induce strong exchange interaction in the surface states via the proximity effect\cite{Jiang2015,Wei2013,Jiang2015_2,Jiang2016}. This opens a way to combine the existing technologies in a S/FI/S Josephson junction on a 3D topological insulator (S/FI-TI/S). Here we theoretically investigate the magnetization reversal processes in such a system.

In this paper we demonstrate that the magnetization of a ferromagnetic interlayer can be successfully reversed by electric current pulse with the amplitude lower than the critical current of the junction, therefore allowing for very low dissipative manipulation by the magnetic moment. The spin-momentum locking provides very large values of the coupling constant between the Josephson and magnetic subsystems. Although the anomalous ground state phase shift occurs in the considered system, we show that the underlying mechanism providing the reversal is the spin orbit torque (SOT)\cite{Yokoyama2010,Yokoyama2011,Mahfouzi2012,Machon2019}, which is caused by the electric current flowing through the junction. We also demonstrate that the theoretical approach used in Refs.~\onlinecite{Konschelle2009,Shukrinov2017,Nashaat2019,Guarcello2020} to describe the magnetization dynamics and reversal in Josephson junctions with metallic ferromagnets, is not applicable to the case of insulating ferromagnets considered here. The detailed time-resolved picture of the magnetization dynamics taking place in the reversal process is investigated. The influence of the magnetic anisotropy of the ferromagnet on the duration and stability of the reversal is shown.

The paper is organized as follows. Sec.~\ref{dynamics} describes the model we study and our theoretical approach to calculating the magnetization dynamics caused by the supercurrent. In Sec.~\ref{sec:reversal} our numerical results on the detailed magnetization dynamics of the reversal process together with the analytical analysis of the influence of the anisotropy are presented. A method of electrical control of the magnetization reversal is also discussed in this section. Our conclusions are summarized in Sec.~\ref{conclusions}.

\section{Magnetization dynamics generated by a Josephson current in S/F/S junction on a 3D TI.}

\label{dynamics}

\subsection{Model system.}

\begin{figure}[!tbh]
   \centerline{\includegraphics[clip=true,width=2.8in]{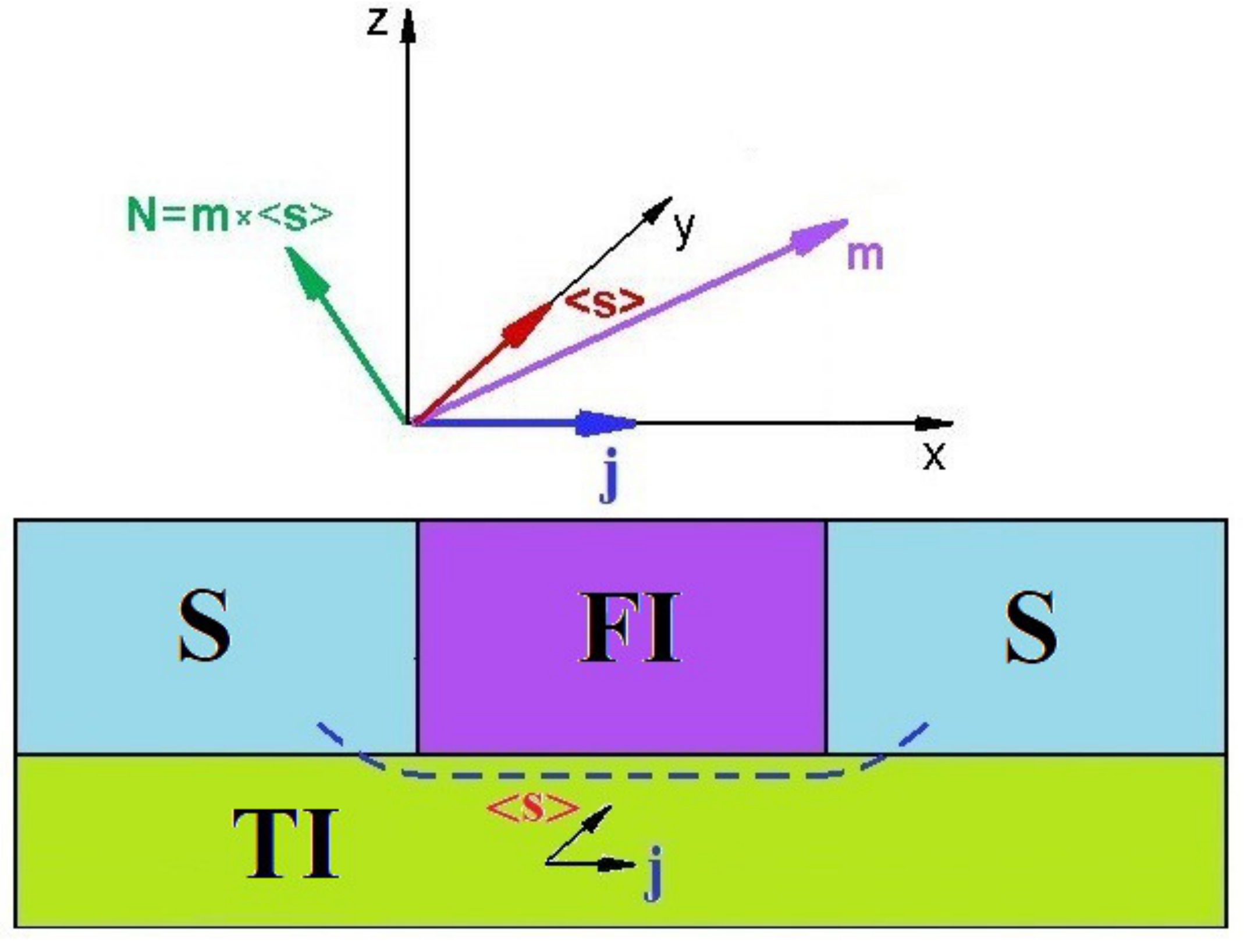}}
        \caption{Sketch of the system under consideration. Superconducting leads and a ferromagnetic insulator as an interlayer are deposited on top of the TI insulator. The dashed line represents a schematic trajectory of the current flow. The induced electron spin polarization $\langle \bm s \rangle$ lies in the TI surface plane and is perpendicular to the current.}
 \label{sketch}
 \end{figure}

The sketch of the system under consideration is presented in Fig.~\ref{sketch}. Two conventional s-wave superconductors and a ferromagnetic insulator (FI) are deposited on top of a 3D TI insulator to form a Josephson junction. The magnetization of the ferromagnet is assumed to be spatially homogeneous. For an insulating ferromagnet all the current flows via the TI surface states. We believe that our results can be of potential interest for systems based on $Be_2Se_3/YIG$ or $Be_2Se_3/EuS$ hybrids, which have been realized experimentally \cite{Jiang2015,Wei2013,Jiang2015_2,Jiang2016}. 

The Hamiltonian that describes the TI surface states in the
presence of the effective exchange interaction between the spin densities on the two sides of the S/F interface reads:

\begin{eqnarray}
\hat H = \hat H_{TI} + \hat H_{int},~~~~~~~~~~~~ \\
\hat H_{TI}=\int d^2 \bm r \hat \Psi^\dagger (\bm r)\Bigl[-iv_F (\bm \nabla \times \bm e_z)\hat {\bm \sigma}-\mu \Bigr]\hat \Psi(\bm r)
\label{H_TI}, \\
H_{int} = - \frac{1}{2}\int d^2 \bm r \hat \Psi^\dagger (\bm r) J_{ex} \bm S \bm \sigma \hat \Psi(\bm r),~~~~~~~~
\label{interface_ham}
\end{eqnarray}
where $\hat \Psi=(\Psi_\uparrow, \Psi_\downarrow)^T$, $v_F$ is
the Fermi velocity, $\bm e_z$ is a unit vector normal to the surface
of TI, $\mu$ is the chemical potential and $\hat {\bm \sigma}=(\sigma_x, \sigma_y, \sigma_z)$ is a vector of
Pauli matrices in the spin space. Here $\bm S$ is the localized spin operator in the FI film, $J_{ex}$ is the exchange constant and the integration is performed over the 2D interface. Eq.~(\ref{interface_ham}) can be written in terms of the effective exchange field $\bm h_{TI} = -(1/2)J_{ex} \bm S$, which is induced by the FI in the TI surface states.

Due to the property of the full spin-momentum locking of the TI surface states an electric current $\bm j$ flowing via the TI surface states induces a stationary electron spin polarization (direct magnetoelectric effect), regardless of the fact if it is a normal current or supercurrent\cite{Shiomi2014,Bobkova2016}
\begin{eqnarray}
\langle \bm s \rangle = -\frac{1}{2ev_F}[\bm e_z \times \bm j].
\label{spin_current}
\end{eqnarray}
This electron spin polarization is coupled to the FI magnetization via the interface exchange term Eq.~(\ref{interface_ham}) giving rise to a torque, acting on the FI magnetization $\bm M = -\gamma \bm S$, where $\gamma$ is the gyromagnetic ratio.

\subsection{Magnetization dynamics}

In the present subsection we formulate the main equations describing  magnetization dynamics of the ferromagnet placed in the interlayer region of the S/3D TI/S Josephson junction. The dynamics is described by the Landau-Lifshitz-Gilbert (LLG) equation. From the exchange interaction Eq.~(\ref{interface_ham}) one obtains the additional contribution to the LLG equation  in the form of a torque acting on the magnetization \cite{Yokoyama2011}:
\begin{eqnarray}
\frac{\partial\bm M}{\partial t} = -\gamma \bm M \times \bm H_{eff} + \frac{\alpha}{M} \bm M \times \frac{\partial\bm M}{\partial t} + \frac{J_{ex}}{d_F} \bm M \times \langle \bm s \rangle,~~~~~~
\label{LLG}
\end{eqnarray}
where $M$ is the saturation magnetization, $\bm H_{eff}$ is the local effective field in the ferromagnet and $\alpha$ is the Gilbert damping constant. The last term represents the torque, averaged over the ferromagnet thickness $d_F$ along the $z$-direction. Taking into account Eq.~(\ref{spin_current}) and assuming that the current is flowing along the $x$-direction, the expression for the torque can be rewritten as
\begin{eqnarray}
\bm N = \frac{J_{ex}}{d_F} \bm M \times \langle \bm s \rangle = -\frac{\gamma h_{TI}j}{eMv_F d_F}[\bm m \times \bm e_y], 
\label{torque}
\end{eqnarray}
where we have introduced $h_{TI} = |\bm h_{TI}|$ and the unit vector $\bm m = \bm M/M$.

The ferromagnet is assumed to be an easy-plane magnet with the hard axis directed along the $z$-axis. An in-plane uniaxial anisotropy along $x$-axis is also assumed. This situation corresponds to the experimental data reported for YIG thin films\cite{Mendil2019}. In this case the local effective field in the ferromagnet can be written as follows:
\begin{eqnarray}
\bm H_{eff} = -\frac{K}{M}m_z \bm e_z + \frac{K_u}{M}m_x \bm e_x, \label{effective_field}
\end{eqnarray}
where $K$ and $K_u$ are hard axis and easy axis anisotropy constants, respectively.

An alternative approach to finding the torque acting on the magnetization in Josephson junctions has been used in Refs.~\onlinecite{Konschelle2009,Shukrinov2017,Nashaat2019,Guarcello2020}. In these papers only the supercurrent-induced part of the torque has been taken into account via the additional contribution to the effective field $\delta \bm H_{eff}$ according to the relation $\delta \bm H_{eff} = -(1/V_F)\delta E_J/\delta \bm M$. Here $V_F$ is the ferromagnet volume and $E_J = (\Phi_0 I_c/2\pi)[1-\cos(\chi-\chi_0)]$ is the Josephson energy, where $\Phi_0$ is the flux quantum, $\chi$ is the phase difference between the superconducting leads and $\chi_0$ is the anomalous ground state phase shift at the junction. In the presence of SOC $\chi_0$ depends on the magnetization direction providing a contribution to $\bm H_{eff}$. This approach is applicable to complex interlayers consisting of metallic ferromagnets and SOC materials or TI, but not suitable for interlayers with insulating ferromagnets. The reason is the following. The current applied to the junction depends on time due to the magnetization dynamics and also if the current is applied in the pulse regime. Then a voltage appears at the junction. The voltage can be found as $V = \dot \chi/2e$ from the following equation, representing the standard RSJJ-model generalized for the presence of the time-dependent anomalous phase shift \cite{Rabinovich2019,Rabinovich2020}:
\begin{eqnarray}
j=j_c \sin (\chi - \chi_0)+\frac{1}{2eR_N}(\dot \chi - \dot \chi_0).
\label{current_total}
\end{eqnarray}
This equation expresses the fact that in the presence of magnetization dynamics the full current $j$ applied to the system is inevitably a sum of the supercurrent contribution $j_s = j_c \sin (\chi - \chi_0)$ and a quasiparticle contribution even if $j<j_c$. Therefore, a part of the full electric current flows as a quasiparticle current.  This current also makes a contribution to the torque expressed by Eq.~(\ref{torque}) on equal footing with the supercurrent. In the case of a metallic ferromagnet the quasiparticle current flows via the ferromagnet, and not through the TI surface states (due to the fact that their resistance is typically much larger) and does not make a torque on the magnetization. However, for the case of insulating ferromagnet the full electric current flows via the TI surface states and is to be taken into account in the torque calculation.   

\section{Magnetization reversal and electrical detection of the reversal}

\label{sec:reversal}

\subsection{Magnetization reversal}

Here we consider numerically the magnetization reversal by the electric current pulse. The LLG equation (\ref{LLG}) in the dimensionless form can be written as follows:
\begin{eqnarray}
\frac{\partial\bm m}{\partial \tilde t} = -\bm m \times \tilde {\bm H}_{eff} + \alpha \bm m \times \frac{\partial\bm m}{\partial \tilde t} + N \tilde j [\bm e_y \times \bm m],~~~~~~
\label{LLG_dimensionless}
\end{eqnarray}
where we have introduced the dimensionless quantities $\tilde t = \gamma t K_u/M$, $\tilde {\bm H}_{eff} = -k m_z \bm e_z + m_x \bm e_x$ with $k=K/K_u$, $\tilde j = j/j_{c0}$ and $ N = rE_J/8E_M$, which is a product of the dimensionless parameter $r=2h_{TI}d/v_F$, quantifying the strength of the SOC, and the ratio of the Josephson $E_J = j_{c0} \Phi_0/2\pi$ and magnetic $E_M = K_u d_F d/2$ energies. Here $j_{c0}$ is the critical current density at $m_x=0$. The parameter $N$ is a measure of the SOT efficiency per unit current. It is worth noting that for the 3D TI-based system the SOT is maximal for a given effective exchange field:  $r$ has a very large value as compared to  spin-orbit coupled materials because it does not contain the reducing factor $\Delta_{so}/\varepsilon_F$, which is the ratio of the spin-orbit splitting to the Fermi energy. The physical reason is the spin-momentum locking of the 3D TI surface states.

\begin{figure}[!tbh]
   \centerline{\includegraphics[clip=true,width=3.0in]{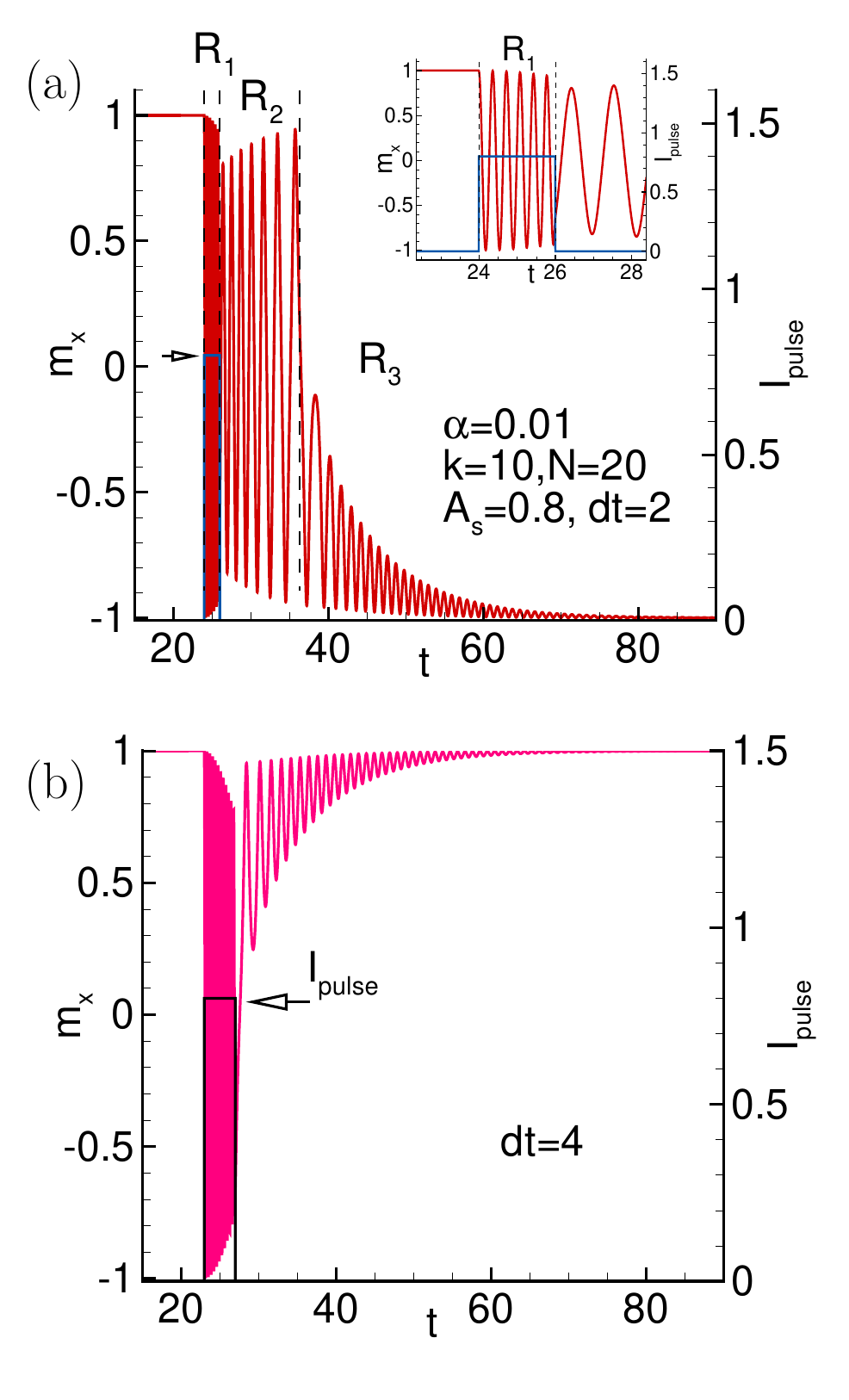}}
        \caption{$m_x(t)$ together with $j(t)$ for two different pulse durations $dt$. The other parameters are the same for the both panels and are given in panel (a). We denote the ratio of the hard-axis and easy-axis anisotropies as $k=K/K_u$. It is seen that the magnetization dynamics consists of three different regimes. They are denoted as $R_1$, $R_2$ and $R_3$ in panel (a). The regime $R_1$ corresponds to the dynamics during the current pulse. It is shown on the large scale in the insert to panel (a). Regime $R_2$ differs from  $R_3$ by the fact that the final state of the magnetization is not determined yet in this regime. Regime $R_2$ is not necessary realized in the system. Such an example is illustrated in panel (b). For a detailed discussion of  regimes $R_{1,2,3}$ see text. }
 \label{reversal}
 \end{figure}
 
Further basing on Eq.~(\ref{LLG_dimensionless}) we investigate the dynamics of the magnetization $\bm$ numerically. For our numerical simulations the parameters of the Josephson junction are taken to be corresponding to $Nb/Bi_2Te_3/Nb$ Josephson junctions \cite{Veldhorst2012}: $d=50 nm$, $j_{c0} = 40 A/m$, $v_F = 10^5 m/s$. Magnetic parameters correspond to YIG thin films \cite{Mendil2019} 
$K_u = 0.5Oe$ and $d_F = 10 nm$. It is difficult to give an accurate a-priori estimate of $h_{TI}$ because there are no reliable experimental data on its value. However, basing on the experimental data on the Curie temperature of the magnetized TI surface states \cite{Jiang2015_2}, where the Curie temperature in the range $20-150K$ was reported, we can roughly estimate $h_{TI} \sim 0.01- 0.1 h_{YIG}$. We assume $h_{TI} \sim 100 K$ in our numerical simulations. Using the above data we obtain $r = 13.2$ and $N = 21.8$. Therefore, the numerical results are calculated at $N=20$.

\begin{figure}[!tbh]
   \centerline{\includegraphics[clip=true,width=3.2in]{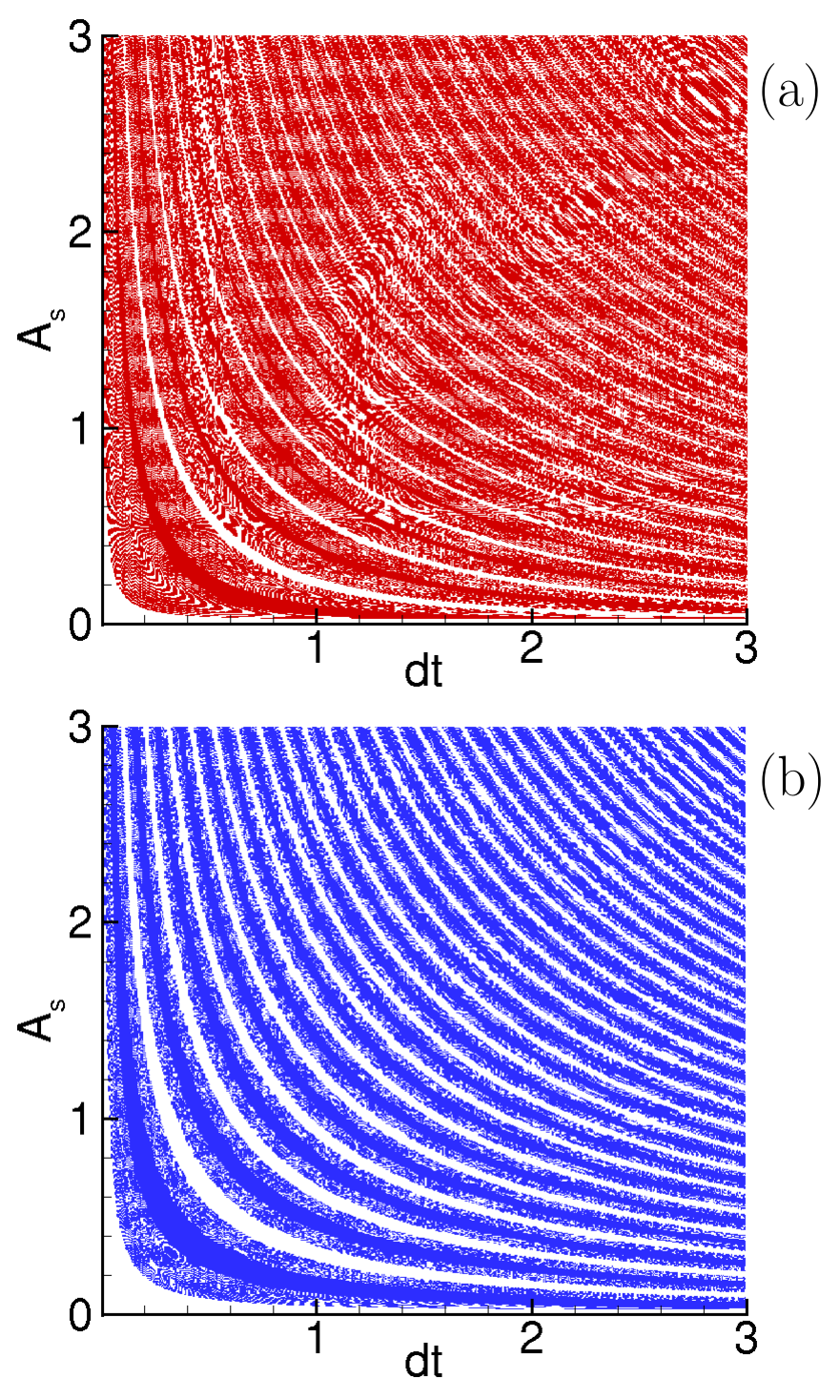}}
        \caption{Yes/no reversal diagram in the $(dt, A_s)$-plane. The regions, where the reversal occurs are colored and where it does not occur are white. They are separated by the white/colored striped regions, which represent an "uncertainty" regime and and discussed in the text. (a) $k=10$, (b) $k=1$. The other parameters are the same as in Fig.~\ref{reversal}.}
 \label{reversal_diagram}
 \end{figure}

To induce the magnetization reversal we apply a rectangular pulse of the electric current with  amplitude $A_s$ and duration $dt$. The resulting magnetization dynamics starting from the initial condition $m_x = 1$ is shown in Fig.~\ref{reversal}. Panels (a) and (b) of this figure correspond to the pulses of the same amplitude but different durations. It is seen that in general the reversal can occur in the system, but if the magnetic moment is reversed depends crucially on the pulse duration. It is illustrated further by numerical data represented in Fig.~\ref{reversal_diagram}. This diagram shows regions, where the reversal occurs/does not occur in the $(dt, A_s)$-plane. It is seen that the reversal is very sensitive to the pulse parameters and the regions, where the reversal occurs (colored) and does not occur (white) are separated by striped regions, where the behavior of the system is hardly predictable. The detailed explanation of this regime is discussed below. It is also remarkable that the lower boundary for the pulse amplitude $A_s$, which is enough for providing the reversal, is much less than unity. Therefore, the magnetic moment can be reversed by the current pulses with amplitudes much lower that the critical Josephson current.   

In Fig.~\ref{reversal} it is also seen that the induced magnetization dynamics has three different regimes, which are separated by dashed lines in Fig.~\ref{reversal}(a) and denoted as $R_{1}$, $R_{2}$, $R_{3}$. In the regime $R_1$ taking place during the current pulse the magnetization tends to align itself to the equilibrium direction for a {\it constant} electric current $\tilde j=A_s$. This direction lies in the $(x,y)$-plane and makes some angle with the easy $x$-axis due to the additional effective field in the $y$-direction produced by the SOT. In case $\tilde j N >1$ the equilibrium magnetization direction is along the $y$-axis. After the pulse ending two strongly different regimes $R_2$ and $R_3$ occur. They are discussed in detail in the next subsection.  

\subsection{Dependence of the reversal on the magnetic anisotropy}

Let us parametrize the magnetization as $\bm m =(\cos \theta, \sin \theta \cos \varphi, \sin \theta \sin \varphi)$. Then after the current pulse ending the magnetic moment evolution is described by the following equations, which are derived from LLG equation Eq.~(\ref{LLG_dimensionless}):
\begin{eqnarray}
(1+\alpha^2) \dot \theta = -\sin \theta \bigl[\frac{k}{2}  \sin 2\varphi  + \alpha \cos \theta  (1+k \sin^2 \varphi)\bigr] \nonumber \\
(1+\alpha^2) \dot \varphi = \cos \theta  (1+k \sin^2 \varphi) - \alpha \frac{k}{2}  \sin 2\varphi . ~~~~~~~~~
\label{LLG_after_pulse}
\end{eqnarray}

\begin{figure}[!tbh]
   \centerline{\includegraphics[clip=true,width=3.4in]{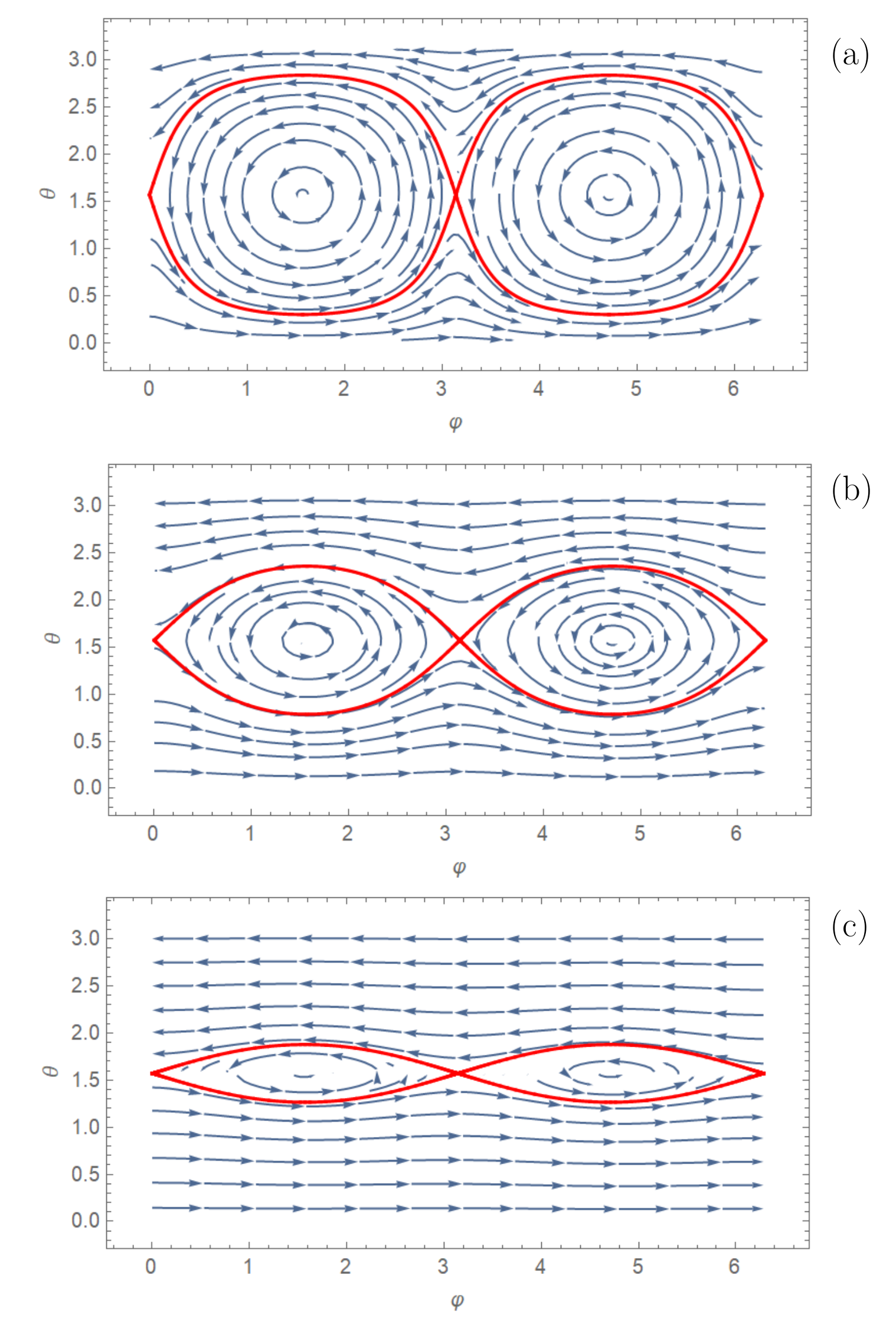}}
        \caption{Vector field $(\dot \varphi, \dot \theta)$ for the free magnetic moment dynamics corresponding to $j=0$ at (a) $k=10$; (b) $k=1$ and (c)$k=0.1$. The Gilbert damping $\alpha=0.01$ for all the panels. The red lines separate uncertainty and stability regions, see text.}
 \label{phase}
 \end{figure}

The vector field $(\dot \varphi, \dot \theta)$ calculated according to Eqs.~(\ref{LLG_after_pulse}) is demonstrated in Fig.~\ref{phase}(a-c) for different values of the anisotropy parameter $k$. There are two different regions in these pictures. If the vector of the magnetization at the moment of the pulse ending gets to the region inside the red boundary, one cannot conclude if it will be reversed after the dynamics decay. On the contrary, if its value at the pulse ending is outside this boundary, one can definitely conclude if it will be reversed. The regions inside and outside the boundary correspond to regimes $R_2$ and $R_3$ in Fig.~\ref{reversal}. While in the regime $R_2$ the magnetization rotates with increasing amplitude around $m_z = \pm 1$, in the regime $R_3$ it tends to one of the equilibrium positions $m_x = \pm 1$. We call the region inside the red boundary by "the uncertainty region" because due to the small value of the Gilbert damping constant the final equilibrium state of the magnetic moment starting from this region is hardly predictable. 

The boundary of the uncertainty region (the red line) can be described analytically. At $\alpha = 0$ system (\ref{LLG_after_pulse}) has the following first integral:
\begin{eqnarray}
I_1 = (1+k\sin^2 \varphi)\sin^2 \theta = const
\label{first_integral}
\end{eqnarray}
It is obvious that $0 \leq I_1 \leq 1+k$. This quantity determines analytically where the reversal definitely happens. To see this let us rewrite Eq.~(\ref{LLG_after_pulse}) at $\alpha = 0$ making use of $I_1$:
\begin{eqnarray}
 \dot \varphi = \pm \sqrt{(1-I_1 + k\sin^2 \varphi)(1+k\sin^2 \varphi)}~~~~~~~~~ \nonumber \\
\dot \theta = -\frac{\sqrt{I_1 - \sin^2 \theta}\sqrt{(k+1)\sin^2 \theta - I_1}}{\sin \theta} . ~~~~~~~~~
\label{LLG_after_pulse_I1}
\end{eqnarray}
We see that there is a critical value of the first integral $I_1^* = 1$. If $I_1^*<I_1<(1+k)$ then to have the quadratic root in $\dot \varphi$ well determined only such values of $\varphi$, which satisfy the condition $1-I_1 + k\sin^2 \varphi>0$ are allowed, what means a finite domain in $\varphi$. Similarly the same idea leads to the finite domain in $\theta$. If we assume $0 \leq I_1 < I_1^*$ we can have arbitrary $\varphi$ and a finite domain in $\theta$. It means that for $0 \leq I_1 < I_1^*$ trajectories in $(\varphi,\theta)$-plane manifest some precession, whereas at $I_1^*<I_1<(1+k)$ they show some finite motion.  These two types of trajectories correspond to the deterministic region outside the red line and to the uncertainty region, respectively. The red line is described by the condition $I_1 = I_1^* = 1$. The first integral Eq.~(\ref{first_integral}) can be rewritten in the other form unveiling its physical meaning: $1-I_1 = \bm m \bm H_{eff} = const$, what simply means that the component of the magnetization along the effective field is conserved during the precession. Then the red line is described by the condition $\bm m \bm H_{eff} = 0$ and the uncertainty region corresponds to $\bm m \bm H_{eff}<0$.

The size of the uncertainty region is determined by the anisotropy parameter $k$, as it can be seen from the above analytical consideration and by comparing panels (a)-(c) of Fig.~\ref{phase}.  The lower the anisotropy parameter the smaller the uncertainty region. In the limit $k \to 0$ the value of the magnetization at the end of the current pulse unambiguously determines if the moment will be reversed, as it has been shown in Ref.~\onlinecite{Mazanik2020}.

\begin{figure}[!tbh]
   \centerline{\includegraphics[clip=true,width=3.4in]{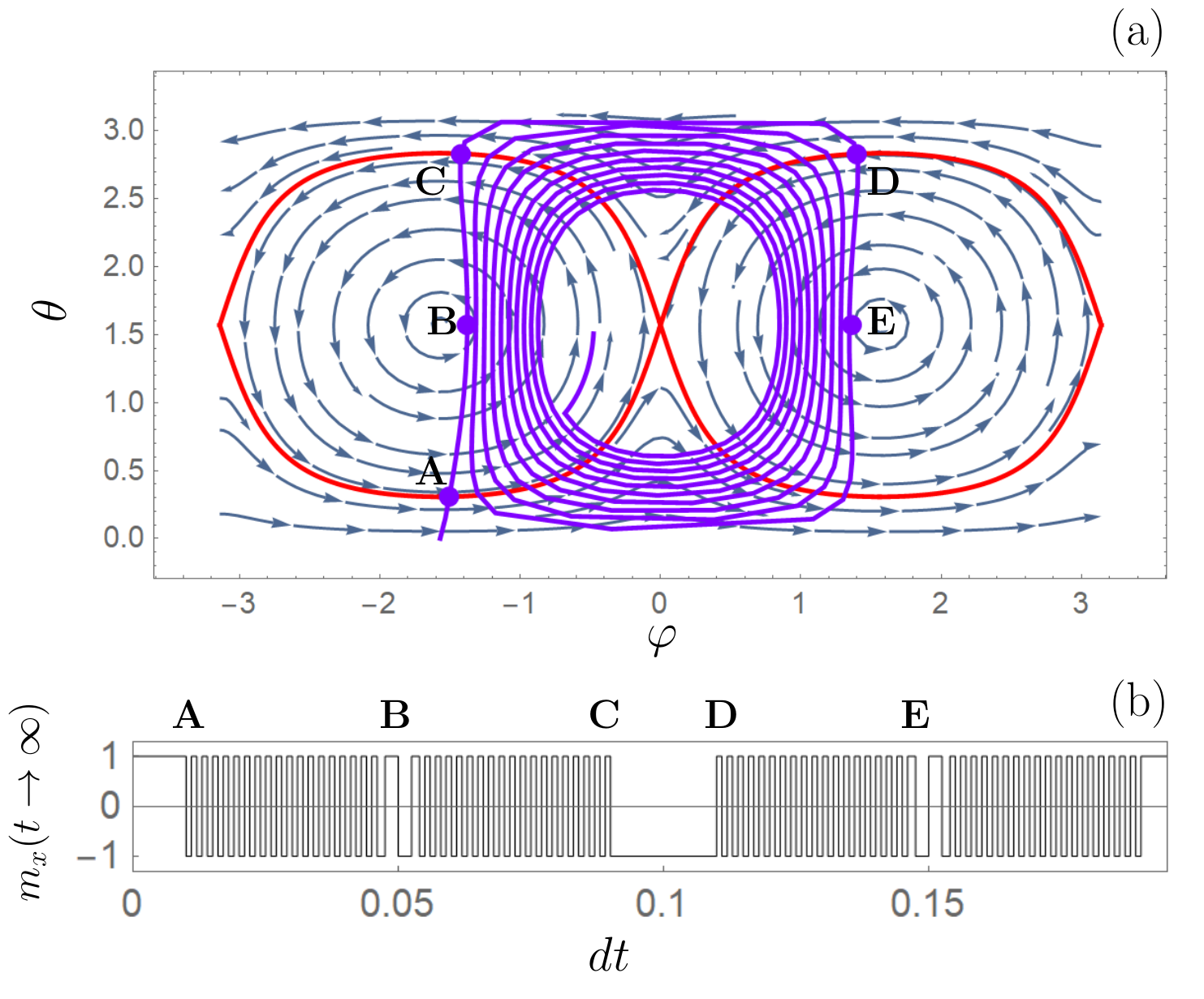}}
        \caption{(a) Trajectory of the magnetic moment evolution during the current pulse (purple curve) put on the vector field $(\dot \varphi, \dot \theta)$ for $k=10$. The pulse can be stopped at any point of the purple trajectory. Some particular moments of the pulse ending are marked by the letters. (b) Magnetic moment final state as a function of the pulse duration $dt$. Letters on top of panel (b) correspond to the same letters in panel (a) matching the particular $dt$ values in panel (b) to the respective position of the magnetic moment at the end of the pulse.}
 \label{mx_dt}
 \end{figure}

Therefore, it can be concluded that large values of the hard-axis anisotropy result in longer times of the reversal process. Also larger values of $k$ lead to lower controllability of the reversal. The point is that in the region outside the uncertainty region the dependence of the final state ($m_x$ is reversed or not reversed) on the pulse duration is simply periodic. However, if the magnetization at the end of the pulse falls into the uncertainty region, the dependence of the final state on the pulse duration becomes strongly oscillating, that is in practice hardly predictable. The illustration of the discussed processes is given in Fig.~\ref{mx_dt}. The purple curve in panel (a) demonstrates the evolution of the magnetic moment during the current pulse starting from $\theta =0, \varphi = -\pi/2$. The moment of the pulse ending corresponds to the break of the curve. Then the magnetic moment evolves according to Eqs.~(\ref{LLG_after_pulse}). Depending on the pulse duration $dt$ the break can occur in the uncertainty region inside the red boundary or in the stability region outside this boundary.  These two possibilities result in very different dependencies of the final state on $dt$. It is shown in panel (b) of Fig.~\ref{mx_dt}. It is seen that if the value of $dt$ is such that the break falls into the uncertainty region, the final state of the magnetic moment is highly oscillating, that is practically unpredictable, as is was stated above. On the contrary, there are wide regions of possible $dt$ values, corresponding to the break outside the uncertainty region, where the final state is determined. Fig.~\ref{mx_dt} is plotted at $k=10$. In this case the width of "unpredictable" regions is rather large and  grows further with enhancing $k$. On the contrary, at small $k$ the width of these regions shrinks and, therefore, the reversal becomes more and more controllable by the current pulses. Fig.~\ref{mx_dt}(b) shows the magnetic moment final state along a line cutting  Fig.~\ref{reversal_diagram} at $A_s=1.5$. Therefore, the oscillating uncertainty regions correspond to the striped white/colored regions of Fig.~\ref{reversal_diagram}.

An analytical criterion for the boundaries of the stability region in $(A_s,dt)$-plane can be easily found at $k \ll N \tilde{j}$ and small $\alpha$, such that $\alpha N dt \ll 1$. From Eq.~(\ref{LLG_dimensionless}) one can find the dynamics of the magnetic moment under the current pulse. Up to the first order with respect to the small parameter $k/\tilde N j$ the solution for the magnetic moment takes the form: 
\begin{equation}
\begin{cases}\label{eq:lowk}
	m^{(0+1)}_x = \cos\left[ N\tilde{j} \tilde{t} \right],\\
	m^{(0+1)}_y = \frac{k+1}{2 N \tilde{j}}\sin^2\left[ N \tilde{j}\tilde{t} \right],\\
	m^{(0+1)}_z = - \sin\left[  N\tilde{j} \tilde{t} \right].
\end{cases}
\end{equation}

Substituting Eq.~(\ref{eq:lowk}) into the criterion of the deterministic region $\bm m(dt) \bm H_{eff}>0$ we obtain the following set of inequalities describing the stability regions in $(A_s,dt)$-plane: 
\begin{equation}
\begin{cases}	\label{eq:newstripes}
	\left( 1 + k  \right)\sin^2[N \tilde{j} d t] < 1,\\
	\frac{\pi}{2} + 2 \pi n	\leq N \tilde{j} d t \leq \frac{3\pi}{2} + 2 \pi n,
\end{cases}
\end{equation}
 The upper of the conditions Eq.~(\ref{eq:newstripes}) corresponds to the falling of $\bm m(dt)$ into the deterministic region outside the red boundary in Fig.~\ref{mx_dt}. The bottom condition follows from the fact that the reversal of the moment corresponds to $m_x(dt)<0$, otherwise the moment returns to its initial value after the decaying of the dynamics.

\subsection{Electrical detection of magnetization reversal}

As it was already mentioned above, the current-induced magnetization dynamics is inevitably accompanied by a finite voltage at the Josephson junction even if the applied current $j$ is stationary and is less than the critical current of the junction. Therefore, the Josephson junction is always in the resistive state in the presence of magnetization dynamics\cite{Rabinovich2019}. The voltage induced at the junction due to the magnetization dynamics can be exploited for an electrical detection of the magnetization reversal.

The voltage $\dot \chi/2e$  at the junction is determined by Eq.~(\ref{current_total}). For the case of the combined F/TI interlayer both the critical current $j_c$ and the anomalous ground state phase shift $\chi_0$ are sensitive to the magnetization dynamics and for the temperatures close to the critical temperature $T_c$ are written as follows\cite{Nashaat2019}:
\begin{eqnarray}
j_c = \frac{ev_F N_F \Delta^2}{\pi^2 T} \int \limits_{-\pi/2}^{\pi/2} d \phi \cos \phi  \times \nonumber \\
\exp[-\frac{2\pi T d}{v_F \cos \phi}] \cos [\frac{2h_{TI,x}d \tan \phi}{v_F}], 
\label{critical_current} \\
\chi_0 = 2 h_{TI,y} d/v_F \label{chi_0}.
\label{anomalous_phase} 
\end{eqnarray}
The quasiparticle current $j_n = (1/2eR_N)(\dot \chi - \dot \chi_0)$ is calculated\cite{Rabinovich2020} in the same approximation $T_c - T \ll T_c$. The resistance of the S/TI/S junction in the normal state $R_N = \pi/e^2 N_F v_F$. It is seen that in the presence of magnetization dynamics there is  an electromotive force ${\cal E} = \dot h_{TI,y} d /(e v_F)$ in the TI resulting from the emergent electric field induced due to the simultaneous presence of the time-dependent exchange field and spin-momentum locking.

\begin{figure}[!tbh]
   \centerline{\includegraphics[clip=true,width=2.8in]{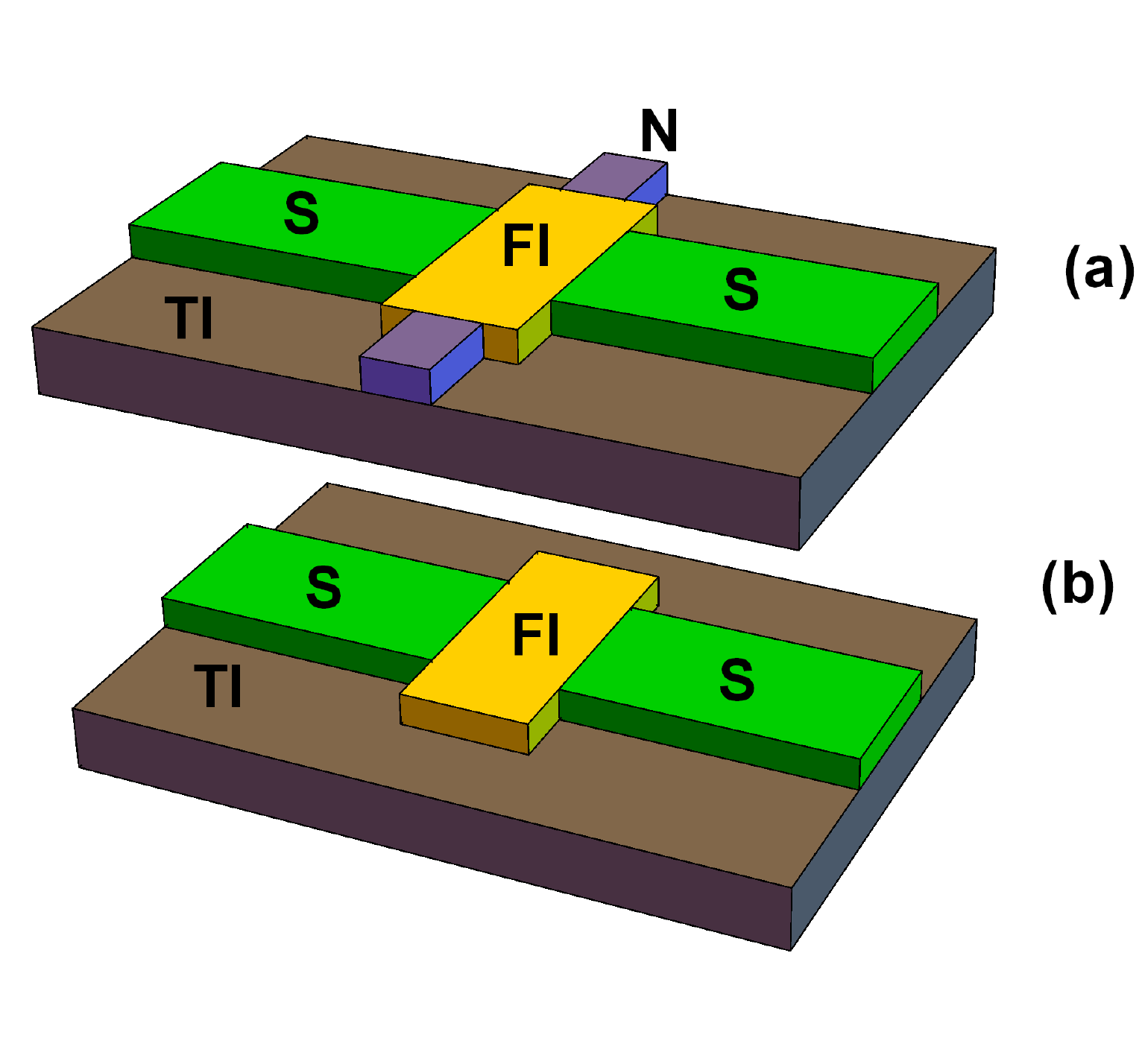}}
        \caption{(a) Sketch of the system with additional normal electrodes, which are used for the electrical detection of $\bm m = \bm e_x \to -\bm e_x$ reversal in comparison to (b) the basic system.}
 \label{sketch_2}
 \end{figure}

If we are interested in the electrical detection of the magnetization reversal $\bm m = \bm e_x \to -\bm e_x$, then it is efficient to measure the transverse voltage generated between the additional leads, as it is shown in Fig.~\ref{sketch_2}. This voltage is measured in the open circuit geometry when the electric current between the additional transverse leads is zero. In this case the solution of Eq.~(\ref{current_total}) takes the form $\dot \chi = \dot \chi_0$ and the voltage generated between the additional leads due to magnetization dynamics can be written as follows\cite{Rabinovich2020}:
\begin{eqnarray}
V_t = \dot h_{TI,x} d/e v_F.
\label{voltage}
\end{eqnarray}
The voltage is determined by the dynamics of $m_x$. It is the same both for superconducting leads and for nonsuperconducting leads and is only determined by the electromotive force. If the magnetization dynamics is caused by the pulse of electric current applied in the $x$-direction, then 
\begin{eqnarray}
\int V_t(t)dt = r  \frac{\hbar}{e} \frac{\Delta m_x}{2},  
\label{voltage_integral}
\end{eqnarray}
where $\Delta m_x$ is the full change of $m_x$ caused by the pulse. If the magnetization reversal $\bm m = \bm e_x \to -\bm e_x$ occurred, then $\Delta m_x = -2$, otherwise it is zero. Therefore, this quantity can be used as a criterion of the magnetization reversal.

\section{Conclusions}

\label{conclusions}

It is shown that the magnetization of an insulating ferromagnet in S/F-3D TI/S can be successfully reversed by an electric current pulse with the amplitude lower than the critical current of the structure. The underlying physical mechanism is the spin-orbit torque. It is demonstrated that the spin-orbit torque is provided by the total current though the junction including both the supercurrent and the quasiparticle current contributions and cannot be calculated in the framework of previously proposed approach, which is based on the calculation of $\varphi_0$-induced additional contribution to the effective field. This approach is only takes into account the supercurrent contribution and is applicable to S/F-3D TI/S junctions with metallic ferromagnets. The influence of magnetic anisotropy on the reversal is investigated. It is found that the presence of strong hard-axis anisotropy reduces the controllability of the reversal making the dependence of the final state on the pulse duration unpredictable. A mechanism of simultaneous electrical detection of the reversal, which is based on the measurement of the voltage generated by the magnetization dynamics, is proposed.

\section*{Acknowledgements}
The reported study was partially funded by the RFBR research projects  18-02-00318 and 18-52-45011-IND.  Numerical calculations were funded by RFBR in the framework of project  number 20-37-70056. The analysis of the role of anisotropy has been supported by RSF project No. 18-72-10135. K.S. thanks DST for support through INT/RUS/RFBR/P-314.

\end{document}